\documentclass[twocolumn]{aastex701}
\usepackage{amsmath}
\usepackage{CJKutf8}

\begin{document}

\title{The Dependence of the Extinction Coefficient on Reddening for Galactic Cepheids}

\author[orcid=0009-0001-2509-5494,sname='Wang']{Huajian Wang}
\affiliation{CAS Key Laboratory of Optical Astronomy, National Astronomical Observatories, Chinese Academy of Sciences, Beijing 100101, China}
\affiliation{School of Astronomy and Space Science, University of the Chinese Academy of Sciences, Beijing, 100049, China}
\email{wanghuajian@bao.ac.cn}  

\author[orcid=0000-0001-7084-0484,sname='Chen']{Xiaodian Chen}
\affiliation{CAS Key Laboratory of Optical Astronomy, National Astronomical Observatories, Chinese Academy of Sciences, Beijing 100101, China}
\affiliation{School of Astronomy and Space Science, University of the Chinese Academy of Sciences, Beijing, 100049, China}
\affiliation{Institute for Frontiers in Astronomy and Astrophysics, Beijing Normal University, Beijing 102206, People’s Republic of China}
\email[show]{chenxiaodian@nao.cas.cn}  

\author[orcid=0000-0003-4489-9794,sname='Wang']{Shu Wang}
\affiliation{CAS Key Laboratory of Optical Astronomy, National Astronomical Observatories, Chinese Academy of Sciences, Beijing 100101, China}
\affiliation{School of Astronomy and Space Science, University of the Chinese Academy of Sciences, Beijing, 100049, China}
\email[show]{shuwang@nao.cas.cn}      

\begin{abstract}

Cepheids are fundamental distance indicators, playing a crucial role not only in the cosmic distance ladder but also in mapping the structure, kinematics, and extinction properties of the Milky Way. Using high-precision photometry and parallaxes from \textit{Gaia} Data
Release 3, we identify a significant anti-correlation between the $G$-band extinction coefficient and reddening for Galactic Cepheids, quantified as $R_G = 1.918 \pm 0.060 - (0.106 \pm 0.022)\,E(G_{\mathrm{BP}} - G_{\mathrm{RP}})$. We propose that this anti-correlation arises from the combination of the non-linear effects inherent to the broad \textit{Gaia} bands and the $R_V$ variations caused by diverse interstellar medium. Adopting a fixed $R_G$ would not only lead to an overestimation of the metallicity dependence of Cepheid luminosities, but also systematically underestimate the distances to highly reddened Cepheids.
Moreover, the strong reddening dependence of $R_G$ makes Wesenheit function based on it unsuitable for highly reddened Cepheids, since the definition of Wesenheit magnitudes requires a fixed extinction coefficient. In contrast, infrared-based distances, being less affected by non-linear effects and insensitive to $R_V$, provide the most reliable Cepheid distances at present.
This work emphasizes the importance of accurately determining $R_V$ for Galactic Cepheids and accounting for non-linear effects in distance measurements, particularly in the optical bands.
\end{abstract}



\section{Introduction}\label{sec:intro}
The period--luminosity (PL) relation for Cepheids \citep{leavitt1912} enables Cepheids to serve as standard candles to trace the structure of the Galactic disk \citep[e.g.,][]{chen2019,skowron2019}, measure distances to nearby galaxies \citep[e.g.,][]{freedman2001,sandage2006}, and determine the late-time expansion rate of the Universe \citep[e.g.,][]{freedman2012,riess2022}. Both the calibration and application of the PL relation critically depend on accurate extinction estimates. Since extinction is difficult to measure directly, it is commonly estimated as the product of reddening and an extinction coefficient. Therefore, accurate determinations of both quantities are essential. The reddening of a Cepheid can be readily derived from the difference between its observed and intrinsic color, where the intrinsic color can be inferred from the period--color (PC) relation \citep[e.g.,][]{Tammann2003}. However, determining the extinction coefficient for each individual Cepheid remains challenging, as it depends on the properties and amount of interstellar dust, the stellar spectra, and the filter transmission curve \citep{maiz2024}.

The properties of interstellar dust, particularly the dust grain size distribution, determine the shape of the extinction curve, $A_\lambda/A_V$ \citep{Draine2001}. Smaller dust grains scatter and absorb shorter-wavelength photons more efficiently, producing steeper extinction curves. Such variations in the extinction curve can be approximately described by a one-parameter family characterized by the ratio of total to selective extinction, $R_V$ \citep{cardelli1989,Wang2023,zhang2025}. Consequently, smaller $R_V$ values correspond to steeper extinction curves and smaller average dust grains.
Previous studies have shown that the $R_V$ of the diffuse interstellar medium (ISM) in the Milky Way is approximately 3.1 \citep[e.g.,][]{savage1979,wang2019}. Therefore, a common simplification is to assume $R_V = 3.1$ universally and adopt one extinction law \citep[e.g.,][]{Fitzpatrick1999,maiz2014,Gordon2023} to derive extinction coefficients for specific bands. This approach is generally acceptable in low-reddening regions dominated by the diffuse ISM, but extrapolating it to highly reddened regions dominated by a denser ISM (e.g., molecular clouds) is questionable. Recent studies have indeed reported significant differences in $R_V$ between molecular clouds and the diffuse ISM \citep[e.g.,][]{maiz2018,Zhang2023,zhang2025}.

Even for a fixed value of $R_V$, the extinction coefficient in a given band is not constant but varies with the amount of dust, owing to a non-linear effect introduced by the finite width of the band. This effect arises from the convolution of the extinction curve with the stellar spectral energy distribution and the filter transmission curve. As extinction increases, shorter-wavelength photons are preferentially absorbed and scattered by dust, shifting the effective wavelength of the band toward longer wavelengths and thereby altering the extinction coefficient \citep[e.g.,][]{Blanco1956,Blanco1957,wang2019,Zhang2023ApJS,maiz2024}. This effect is more pronounced in broader or shorter-wavelength bands.

Benefiting from the high-precision photometry provided by \textit{Gaia} \citep{Gaia2016,Gaia2023}, many recent studies have used the \textit{Gaia} $G$ band to construct and exploit the standard-candle properties of Cepheids (as well as other variable stars). However, using more reliable extinction and distance estimates for Galactic Cepheids, \citet{wang2025} identified that the $G$-band distances systematically deviate from the infrared distances as extinction increases (see their Figure 5). This systematic deviation is also evident in the lower panel of Figure 18 of \citet{Skowron2025}. In this work, we show that this systematic deviation arises from ignoring the strong reddening dependence of $R_G$ (i.e., $A_G / E(G_{\mathrm{BP}} - G_{\mathrm{RP}})$). We propose that this reddening dependence arises from the combined influence of two factors. The first factor is the pronounced non-linear effect inherent to the broad \textit{Gaia} bands \citep[see Appendix B of][]{Green2021}, as discussed in Section \ref{subsec:Convolution Method}. The second factor is the variations in $R_V$ caused by diverse ISM, which is further discussed in Section \ref{subsec:4.2}.

This paper is organized as follows: In Section \ref{sec:data}, we describe the data used in this study. In Section \ref{sec:Result}, we analyze the non-linear effect and the reddening dependence of $R_G$. In Section \ref{sec:discussion}, we present our discussions. Finally, in Section \ref{sec:conclusion}, we summarize our conclusions.

\section{Data} \label{sec:data}

We consider two samples. The first, hereafter referred to as the \textit{Gaia} sample, comprises 1,002 Cepheids with spectroscopic metallicities drawn from \cite{Trentin2024jan} and \cite{Trentin2024oct}, where the former compiles metallicity measurements from multiple literature sources \citep[e.g.,][]{Groenewegen2018,Ripepi2021,Kovtyukh2022,Recio2023,Trentin2023}. We retain only Cepheids with $\mathrm{RUWE} < 1.4$ and apply the parallax correction following \citet{Lindegren2021}. The second sample, hereafter the infrared sample, is taken from \cite{wang2025}, who calibrated and applied period--luminosity--metallicity (PLZ) relations in the infrared bands ($J$, $H$, $K_S$, $W1$, $W2$, [3.6], and [4.5]) and derived the most accurate distances to date for 3,452 Cepheids from \citet{Pietrukowicz2021} using a multi-band optimal distance method \citep[e.g.,][]{Freedman1991,Chen2018}.

For both samples, we retain only Cepheids pulsating in the fundamental (F), first-overtone (1O), or multi-mode (F1O), converting 1O periods to their F-mode equivalents using the updated empirical relation from \citet{Skowron2025}. We then retrieve intensity-averaged magnitudes ($m_{G}$, $m_{G_\mathrm{BP}}$, and $m_{G_\mathrm{RP}}$) from the gaiadr3.vari\_cepheid catalog \citep{Ripepi2023} and retain only Cepheids with reliable photometry \citep[$m_{G_{\mathrm{BP}}} < 20.3$ mag; see Section 8.1 of][]{Riello2021}. The final \textit{Gaia} and infrared samples contain 893 and 2,846 Cepheids, respectively.

\begin{figure}[ht!]
\includegraphics[width=0.45\textwidth]{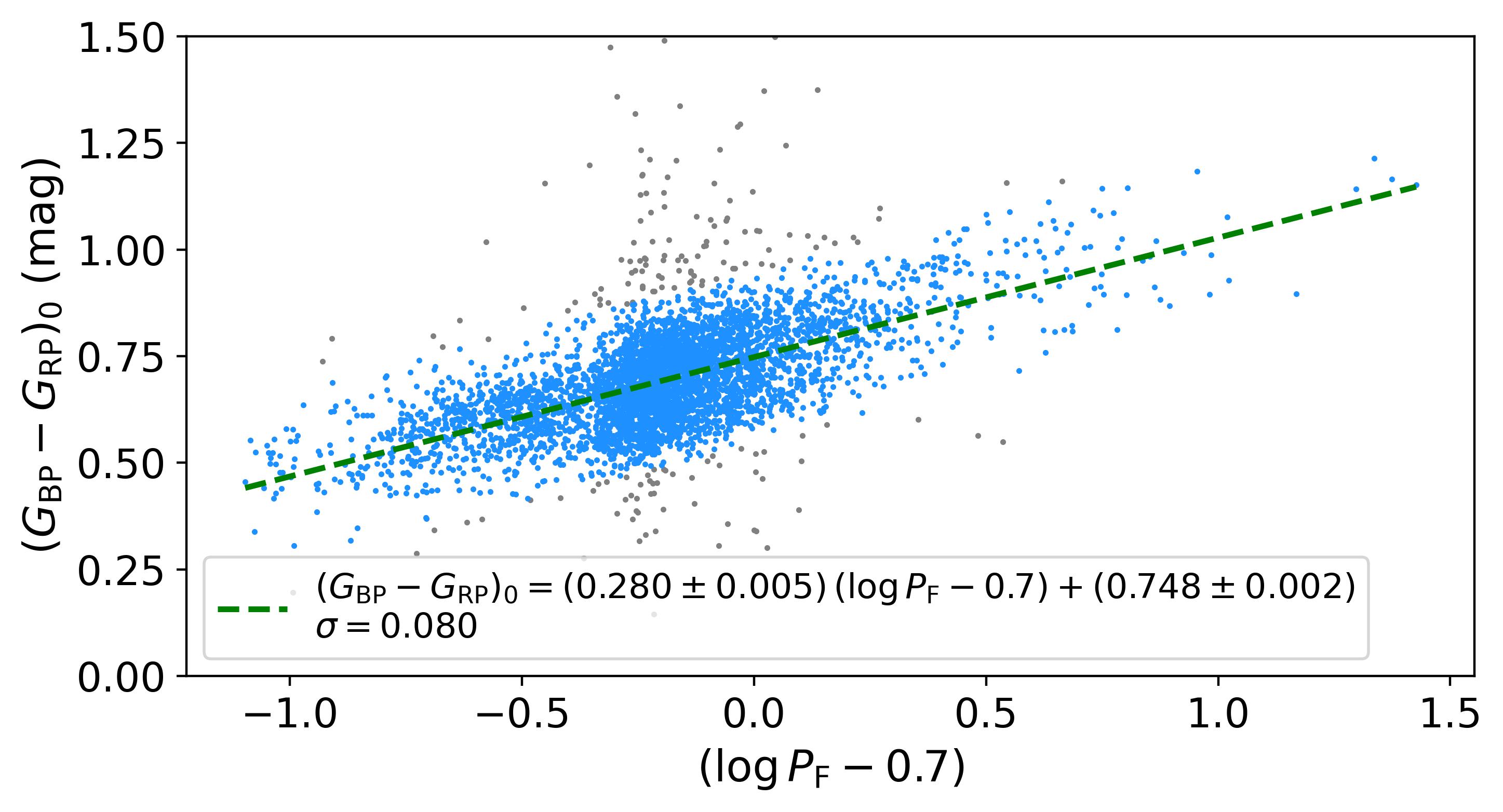}
\caption{The blue dots represent the 4120 LMC Cepheids used for the fit, while the gray dots indicate the 174 outliers rejected through iterative clipping. The green dashed line shows the fitted PC relation, with a standard deviation of 0.08 mag.}
\label{fig:10}
\end{figure}

The reddening $E(G_{\mathrm{BP}} - G_{\mathrm{RP}})$ for each Cepheid is calculated as
$E(G_{\mathrm{BP}} - G_{\mathrm{RP}}) = (m_{G_\mathrm{BP}} - m_{G_\mathrm{RP}}) - (G_\mathrm{BP}-G_\mathrm{RP})_0$. To determine the intrinsic color $(G_\mathrm{BP}-G_\mathrm{RP})_0$, we adopt a sample of Cepheids in the Large Magellanic Cloud (LMC) from the Optical Gravitational Lensing Experiment IV survey \citep{Udalski2018}. The photometric data are likewise taken from the gaiadr3.vari\_cepheid catalog. Extinction is derived using the LMC reddening map of \citet{Skowron2021}, combined with  the LMC extinction law of \citet{Wang2023}. The fitting procedure employs iterative $3\sigma$ clipping to remove outliers, and the results are shown in Figure \ref{fig:10}. The resulting PC relation is
$(G_\mathrm{BP}-G_\mathrm{RP})_0 = (0.280 \pm 0.005)(\log P_{\textrm{F}} - 0.7) + (0.748 \pm 0.002)$. Our result is consistent with the PC relation obtained by subtracting the $G_\mathrm{BP}$ and $G_\mathrm{RP}$ PL relations from \citet{Breuval2022}, providing independent validation. The standard deviation of our PC relation is approximately 0.08 mag. After accounting for the photometric uncertainties in $m_{G_\mathrm{BP}}$ and $m_{G_\mathrm{BP}}$, we estimate an intrinsic scatter of $(G_\mathrm{BP}-G_\mathrm{RP})_0$ of about 0.078 mag. We derive the PC relation from the LMC mainly because the reddening there is very low (with an average $E(G_{\mathrm{BP}} - G_{\mathrm{RP}}) \approx 0.133$ mag), which greatly reduces errors caused by reddening. Moreover, Cepheids in the LMC can be considered to lie at a common distance \citep{pie2019}, making it an excellent laboratory for calibrating the PC relation.

\section{Analysis} \label{sec:Result}
\subsection{Tracing the Non-linear Effects via Convolution Method}\label{subsec:Convolution Method}

\begin{figure*}[ht!]
\centering
\begin{minipage}{0.329\textwidth}
\includegraphics[width=\linewidth]{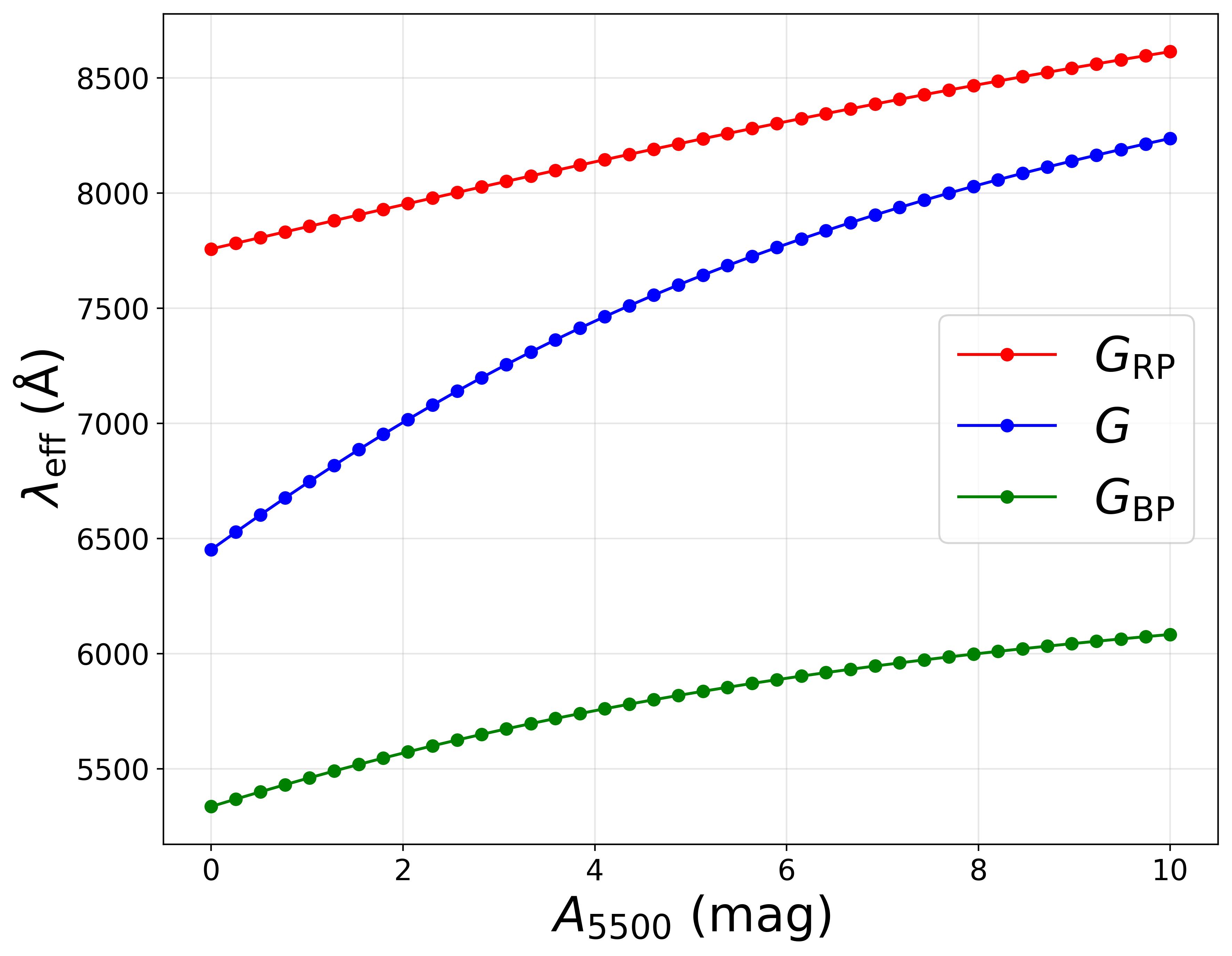}
\end{minipage}
\begin{minipage}{0.329\textwidth}
\includegraphics[width=\linewidth]{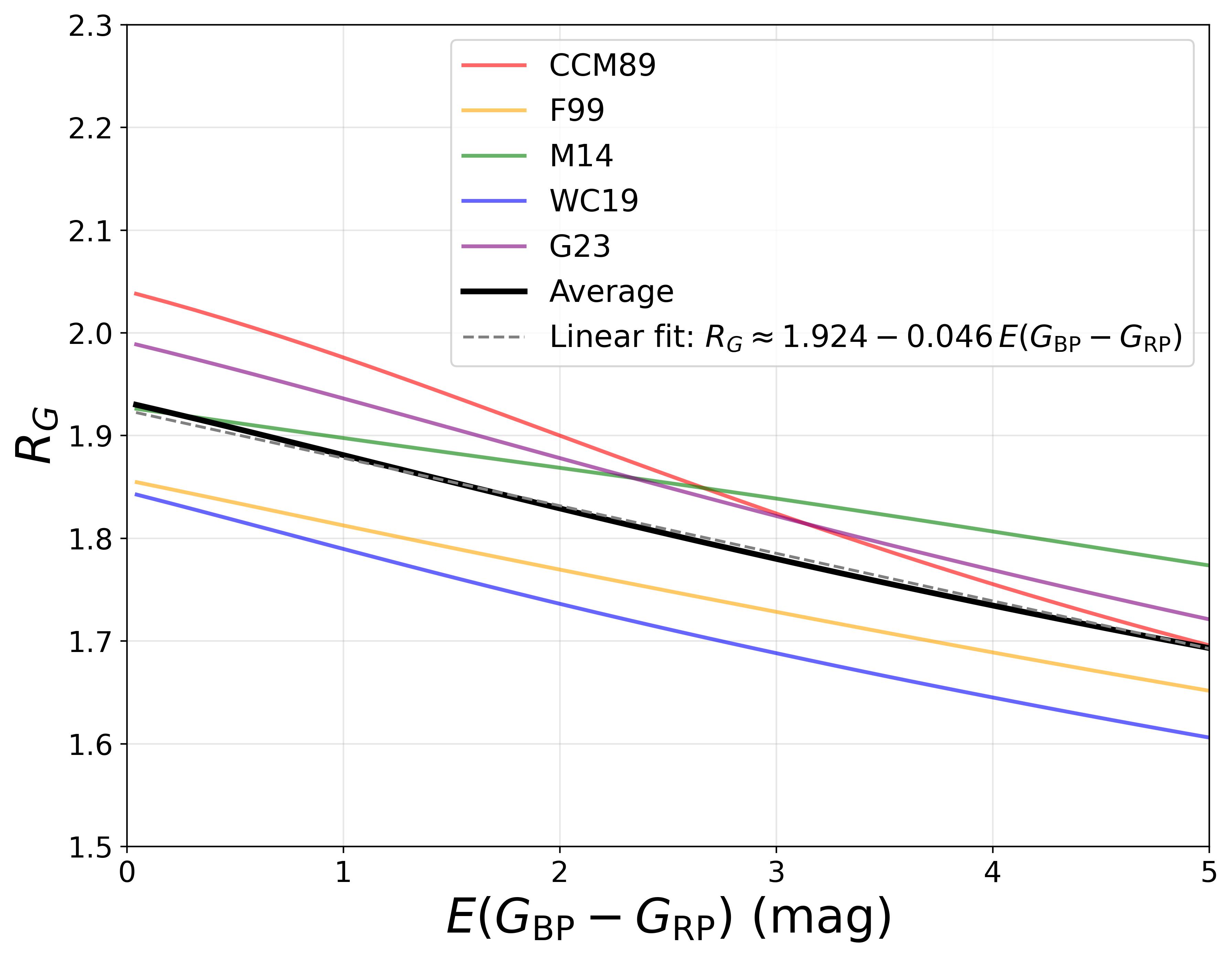}
\end{minipage}
\begin{minipage}{0.329\textwidth}
\includegraphics[width=\linewidth]{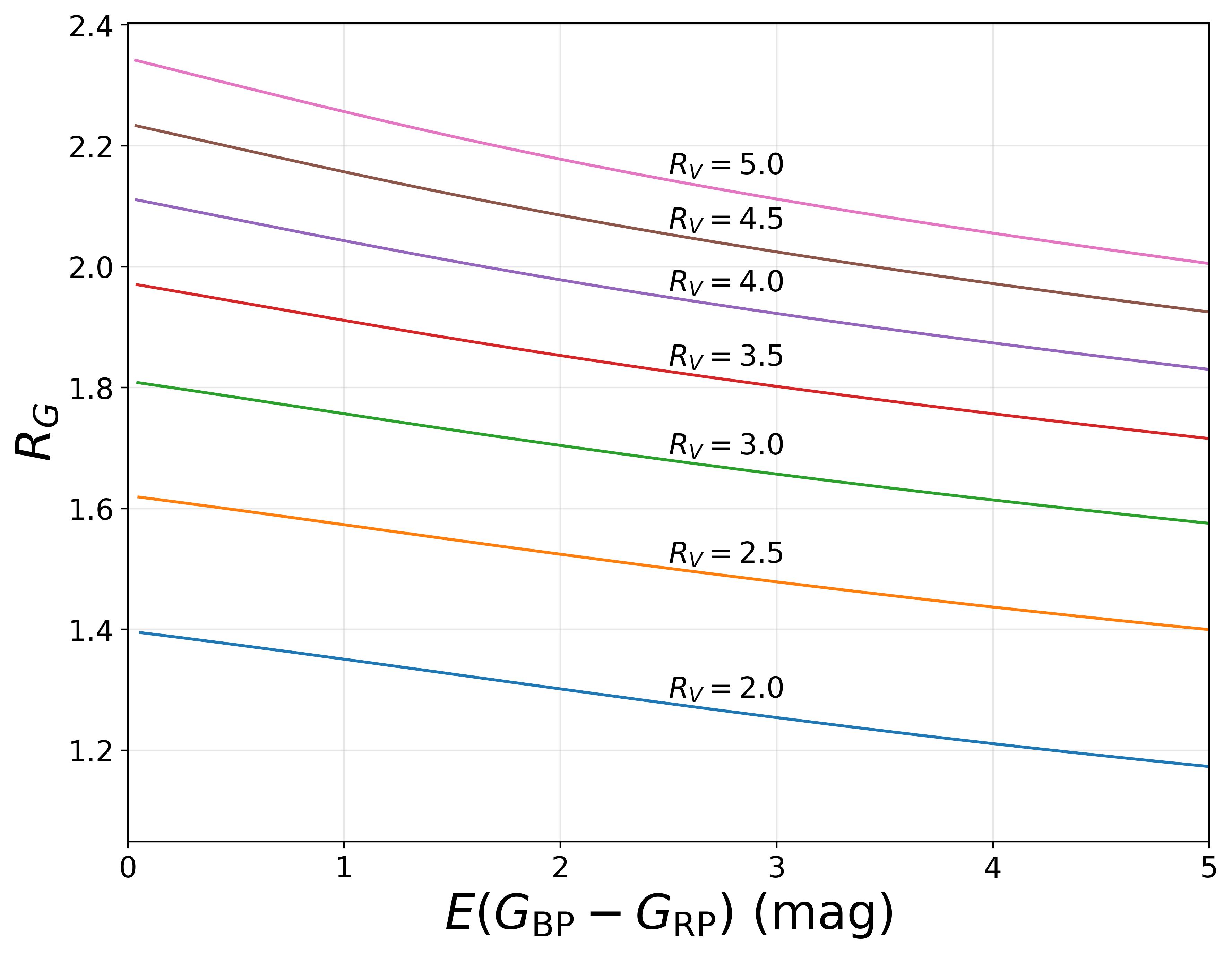}
\end{minipage}
\caption{The left panel shows the non-linear effect of the effective wavelength in the \textit{Gaia} bands. The middle panel shows the non-linear effects of $R_G$ as a function of $E(G_{\mathrm{BP}}-G_{\mathrm{RP}})$ for five different extinction laws with $R_V = 3.1$, where CCM89 represents \citet{cardelli1989}, F99 represents \citet{Fitzpatrick1999}, M14 represents \citet{maiz2014}, WC19 represents \citet{wang2019}, and G23 represents \citet{Gordon2023}. The black curve represents the average of the five curves, while the dashed gray line denotes a linear fit to this average curve. The right panel shows the distribution of $R_G$ as a function of $E(G_{\mathrm{BP}}-G_{\mathrm{RP}})$ for different values of $R_V$.
\label{fig:1}}
\end{figure*}

In this subsection, we investigate the non-linear effects using the convolution method. First, we examine how the non-linear effect manifests in the effective wavelength of the \textit{Gaia} bands (photon-counting bands) to visualize this effect. The effective wavelength of a photon-counting band is computed as:

\begin{equation}
    \lambda_\mathrm{eff} =  
    \frac{\int F_0(\lambda)\, S(\lambda)\, 10^{-0.4 A_\lambda}\, \lambda^2\, d\lambda}
         {\int F_0(\lambda)\, S(\lambda)\, \lambda\, d\lambda},
\end{equation}
where $F_0(\lambda)$ is the intrinsic flux of the stellar spectra, $F_0(\lambda)\,10^{-0.4A_\lambda}$ represents the extincted flux, and $S(\lambda)$ is the filter transmission curve. The wavelength-dependent extinction is expressed as 
$A(\lambda) = \frac{A_\lambda}{A_{5500}}A_{5500}$, where 5500\,\AA{} approximately corresponds to the central wavelength of the $V$ band. We use synthetic stellar spectra from the updated BOSZ
library \citep{Bohlin2017,Bohlin2024}, adopting spectral parameters representative of the mean physical properties of Galactic Cepheids (i.e.,  $T_\mathrm{eff} = 5500\,\mathrm{K}$, $\log g = 2$, and [Fe/H] = 0). The $Gaia$ filter transmission curve $S(\lambda)$ is adopted from \citet{maiz2025}. The extinction curve $A_\lambda/A_{5500}$ follows the extinction law of \citet{wang2019} with
$R_V = 3.1$, and $A_{5500}$ is varied from 0 to 10\,mag in steps of 0.25\,mag. The resulting non-linear effect of the effective wavelength is shown in the left panel of Figure \ref{fig:1}. As $A_{5500}$ increases, the effective wavelengths of the \textit{Gaia} bands all shift toward longer wavelengths, with the \textit{G} band — the broadest among them — exhibiting the strongest non-linear effect.

Next, we examine the non-linear effect of $R_G$. The extinction in a photon-counting band, $A_x$, is defined as:
 
\begin{equation}
A_x = -2.5 \log_{10} \left( 
\frac{\int F_0(\lambda)\, S(\lambda)\, 10^{-0.4 A_\lambda}\, \lambda\, d\lambda}
{\int F_0(\lambda)\, S(\lambda)\, \lambda\, d\lambda}
\right).
\end{equation}
The non-linear effect of $R_G$ is illustrated in the middle panel of Figure \ref{fig:1}. We find that non-linear effects cause $R_G$ to follow an approximately linear relation with $E(G_{\mathrm{BP}} - G_{\mathrm{RP}})$. However, this relation is sensitive to the adopted extinction law, with slopes ranging from about $-0.03$ to $-0.07$ and intercepts from 1.85 to 2.05. By averaging the curves generated from different extinction laws, we derive their mean trend, for which a linear fit yields $R_G \approx 1.924 - 0.046\, E(G_{\mathrm{BP}} - G_{\mathrm{RP}})$. Furthermore, we find that, compared to its sensitivity to the extinction law, the non-linear effect of $R_G$ is largely insensitive to the choice of stellar spectra within the typical parameter range of Galactic Cepheids (i.e., $4500 < T_{\mathrm{eff}} < 7000~\mathrm{K}$, $0.5 < \log g < 3.5~\mathrm{dex}$, and $-1 < [\mathrm{Fe/H}] < 0.5$). The right panel of Figure \ref{fig:1} shows the impact of adopting different $R_V$ in the convolution method on the non-linear effect of $R_G$, using the extinction law of \citet{wang2019}, indicating that $R_G$ is positively correlated with $R_V$.

During the final stages of preparing this manuscript, we became aware of a recent arXiv preprint by \citet{Skowron2026}, who also independently reported a similar non-linear effect of $R_G$ and a similar correlation between $R_G$ and $R_V$ (see their Figure 3).

\subsection{Tracing the Reddening-Dependent $R_G$ via MCMC Fitting} \label{subsec:MCMC}

In this subsection, we employ the Markov Chain Monte Carlo (MCMC) method to perform Bayesian parameter estimation in order to investigate the reddening dependence of $R_G$ using observational data.

First, we use the \textit{Gaia} sample. In parallax space, the predicted parallax is given by
\begin{equation}
\varpi_\mathrm{model} = 10^{-0.2 (m_G - A_G - M_G) + 2},
\end{equation}
where absolute magnitude $M_G$ is defined as
\begin{equation}
M_G = \alpha(\log P_{\textrm{F}} - 1) + \beta + \gamma[\mathrm{Fe/H}],
\end{equation}
and $A_G$ is defined as
\begin{equation}
A_G =
\begin{cases}
\scriptstyle 1.9\,E(G_{\mathrm{BP}} - G_{\mathrm{RP}}), &\scriptstyle \text{Model 1},\\
\scriptstyle (R_1 + R_2\,E(G_{\mathrm{BP}} - G_{\mathrm{RP}}))\,E(G_{\mathrm{BP}} - G_{\mathrm{RP}}), &\scriptstyle \text{Model 2}.
\end{cases}
\end{equation}
Model~1 adopts the conventional assumption of a constant extinction coefficient (i.e., $R_G = 1.9$). 
In contrast, Model~2 introduces free parameters $R_1$ and $R_2$ to capture the approximately linear relation between $R_G$ and $E(G_{\mathrm{BP}} - G_{\mathrm{RP}})$, as revealed by the convolution method. 

We adopt a Gaussian log-likelihood function:
\begin{equation}
\ln \mathcal{L} = 
-\frac{1}{2}\sum_i \left[
\frac{(\varpi_i - zp - \varpi_{\mathrm{model},i})^2}{\sigma_{\mathrm{tot},i}^2}
+ \ln(2\pi\sigma_{\mathrm{tot},i}^2)
\right],
\end{equation}
where $zp$ is the residual parallax offset \citep[e.g.,][]{Riess2021,Reyes2023,Wang2024}. Since the fitting is performed in parallax space, uncertainties from all relevant quantities are propagated.
The total uncertainty of each Cepheid is given by
\begin{equation}
\sigma_{\mathrm{tot},i}^2 = \sigma_{\varpi_i}^2 +
\sum_{x_i} \left( \frac{\partial \varpi_{\mathrm{model},i}}{\partial {x_i}}\sigma_{x_i} \right)^2,
\end{equation}
where $x_i$ includes $m_G$, $m_{G_\mathrm{BP}}$, $m_{G_\mathrm{RP}}$, $[\mathrm{Fe/H}]$, ${M_G}$ and $(G_\mathrm{BP}-G_\mathrm{RP})_0$. To account for the intrinsic scatter of Cepheids, we conservatively adopt $\sigma_{M_G} = 0.2$ mag for the intrinsic
absolute magnitude dispersion \citep{Ripepi2019} and $\sigma_{(G_\mathrm{BP}-G_\mathrm{RP})_0} = 0.08$ mag for the intrinsic color dispersion (see Figure \ref{fig:10}). For the \textit{Gaia} DR3 parallax uncertainties $\sigma_{\varpi_i}$, we increase them conservatively by 10\%. Uniform priors are adopted for all parameters:
\begin{equation}
R_1,\, R_2,\, \alpha,\, \beta,\, \gamma,\, zp \in (-10,\, 10).
\end{equation}

\begin{deluxetable*}{ccccccccccc}[ht!]
\tabletypesize{\small}
\tablecaption{The fitting results of the $G$-band period–luminosity–metallicity relation \label{tab:1}}
\tablehead{
\colhead{Model} & \colhead{$R_1$} & \colhead{$R_2$} & \colhead{$\alpha$} & \colhead{$\beta$} & \colhead{$\gamma$} & \colhead{$zp$} & \colhead{$\chi^2$} & \colhead{RMSE} & \colhead{AIC} & \colhead{BIC}
 \\
\colhead{} & \colhead{} & \colhead{} & \colhead{} & \colhead{(mag)} & \colhead{} & \colhead{(mas)} & \colhead{} & \colhead{} & \colhead{} & \colhead{}  }
\startdata
1 & \nodata & \nodata & 
-2.773$\pm0.035$ & -4.489$\pm0.025$& -0.608$\pm0.050$& 0.016$\pm0.002$ & 1.03 & 0.049 & -3092 & -3073\\
2 & 1.918$\pm0.060$ & -0.106$\pm0.022$ & -2.712$\pm0.035$ & -4.393$\pm0.040$ & -0.396$\pm0.051$& 0.021$\pm0.002$ & 1.00 & 0.041 & -3245 & -3217\\
3 & 1.962$\pm0.017$ & -0.096$\pm0.004$ & -2.642$\pm0.014$ & -4.306$\pm0.017$& -0.440$\pm0.018$& \nodata & \nodata & \nodata & \nodata & \nodata\\
\enddata
\tablecomments{$\alpha$, $\beta$, $\gamma$, and $zp$ are the slope, intercept, metallicity coefficient, and residual parallax offset, respectively.  
$\chi^2$, RMSE, AIC, and BIC are the reduced $\chi^2$ value, root mean square error, Akaike Information Criterion, and Bayesian Information Criterion, respectively. While $\chi^2$ and RMSE quantify the goodness of fit, AIC and BIC assess the trade-off between goodness of fit and model complexity.}
\end{deluxetable*}

\begin{figure*}[ht!]
\centering
\includegraphics[width=0.49\textwidth]{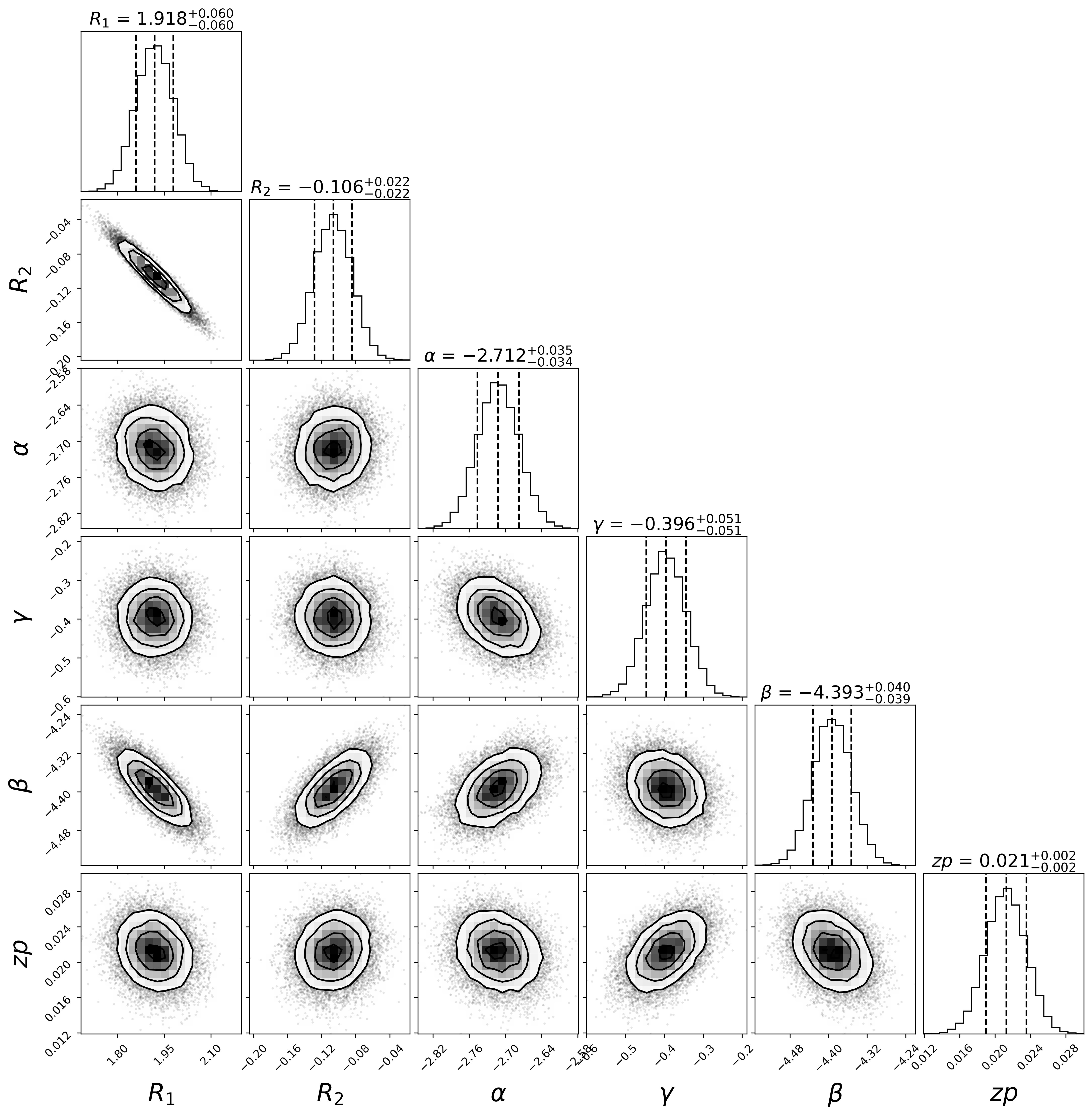}
\hfill
\includegraphics[width=0.49\textwidth]{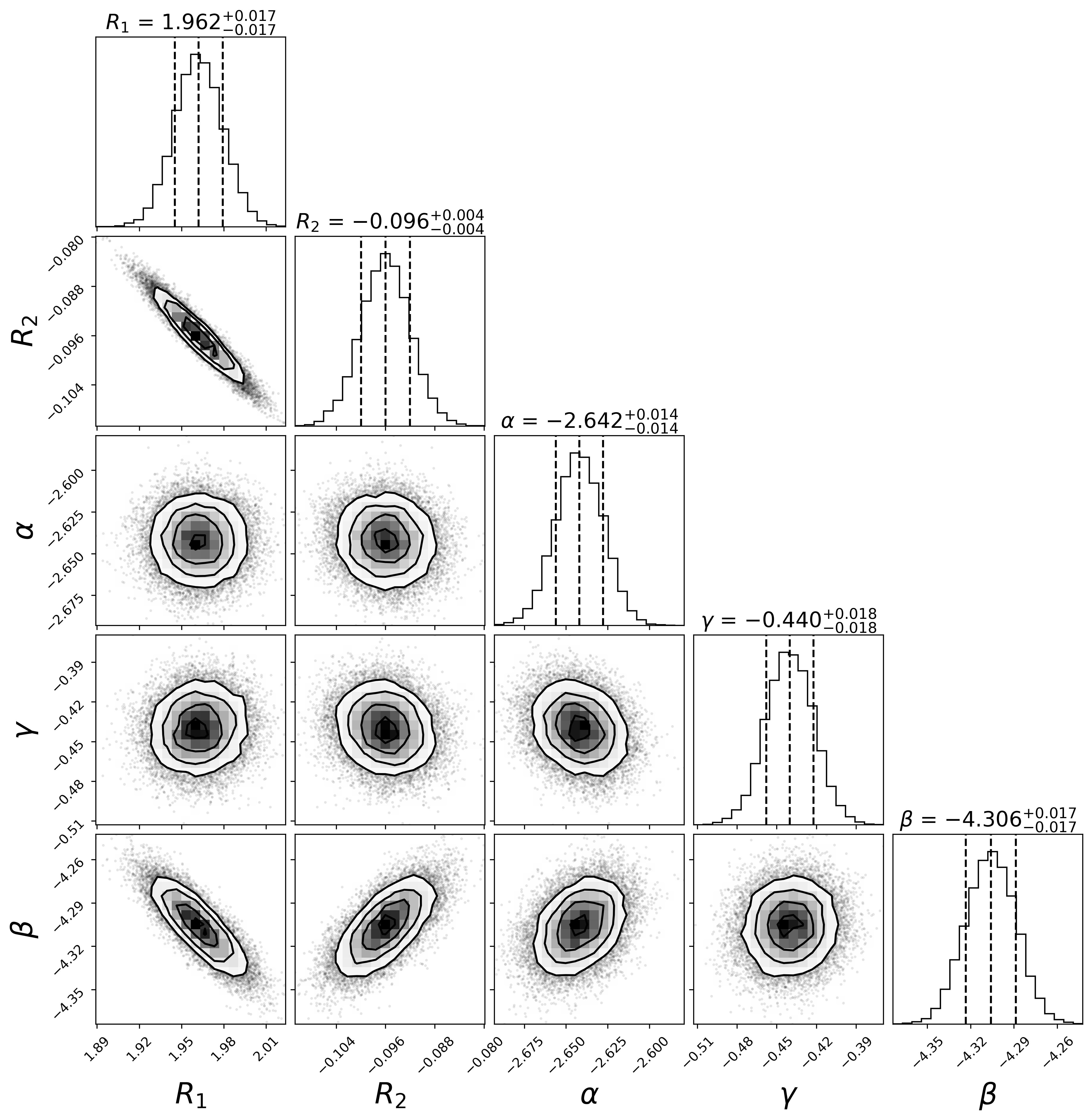}
\caption{The left panel shows the posterior distribution for Model~2, while the right panel shows that for Model~3.
\label{fig:2}}
\end{figure*}

The MCMC sampling is performed using the \texttt{emcee} Python package \citep{emcee2013} with 32 walkers and 10,000 iterations, discarding the first 1,000 steps as burn-in. During the fitting, a $3\sigma$ iterative clipping is applied to remove outliers, resulting in the exclusion of 16 and 13 Cepheids for Model~1 and Model~2, respectively. The median of the posterior distribution is adopted as the best-fit value. The fitting results for the two models are summarized in Table~\ref{tab:1}, and the left panel of Figure~\ref{fig:2} shows the posterior distributions of the Model~2 parameters as an example.

\begin{figure*}[ht!]
\centering
\includegraphics[width=0.49\textwidth]{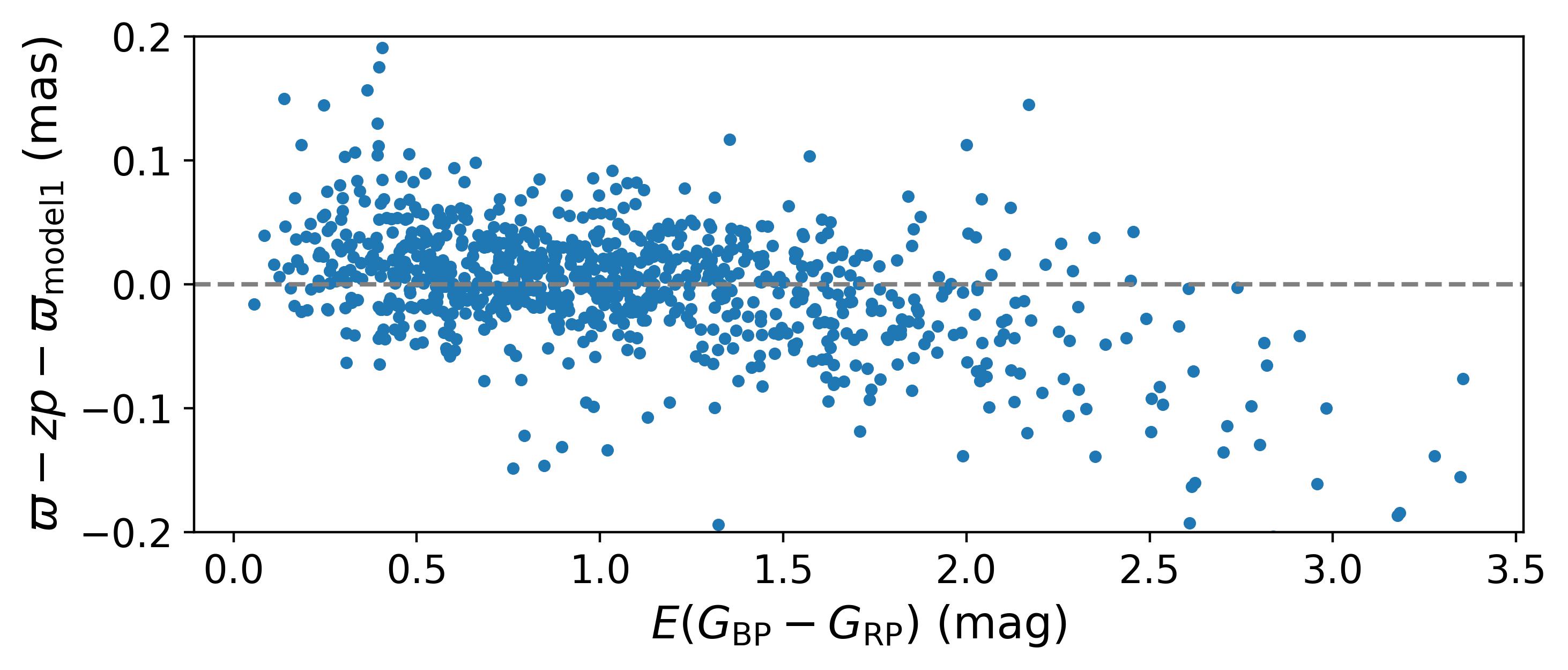}
\hfill
\includegraphics[width=0.49\textwidth]{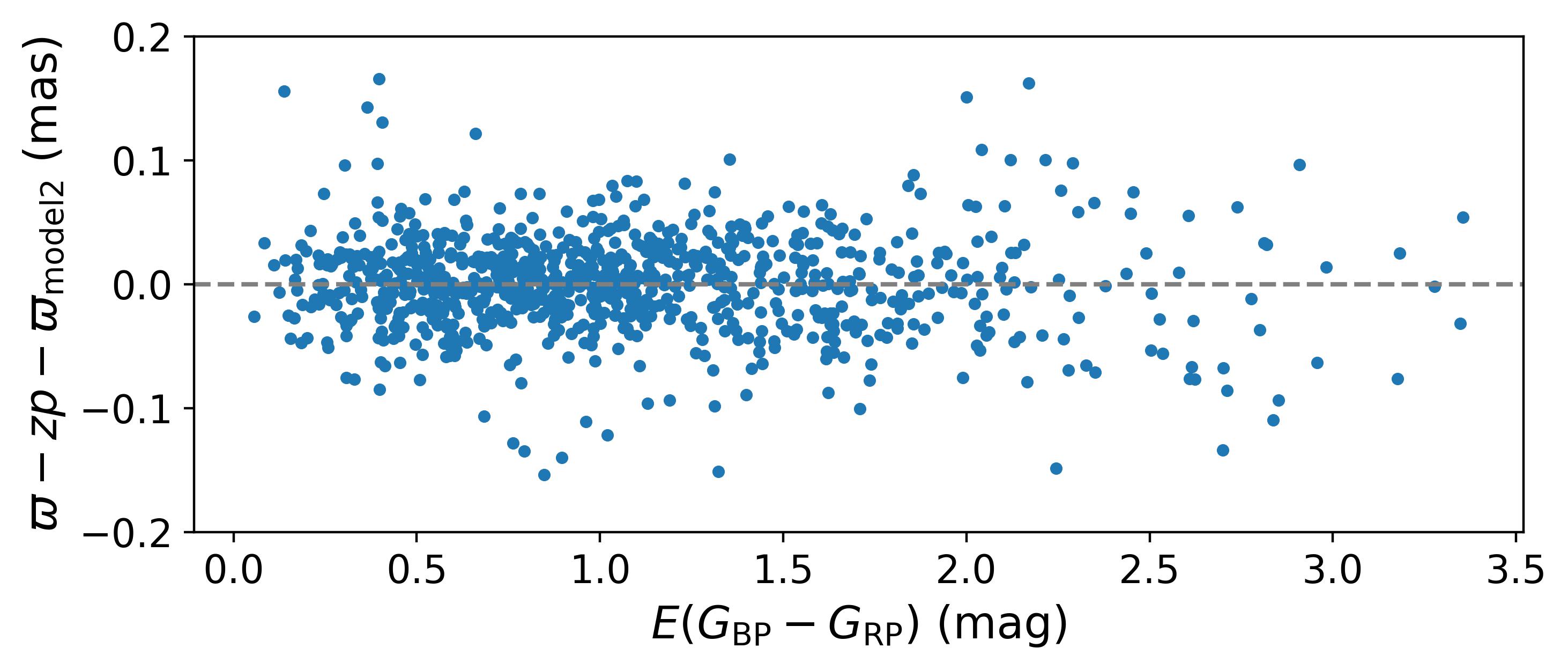}
\caption{The left panel shows the distribution of parallax residuals as a function of
$E(G_{\mathrm{BP}} - G_{\mathrm{RP}})$ for Model~1, while the right panel shows that for Model~2.}
\label{fig:3}
\end{figure*}

Comparing the reduced $\chi^2$, the root-mean-square error (RMSE), and the information criteria (AIC and BIC) listed in Table \ref{tab:1}, Model~2 is statistically preferred over Model~1. Since the comparison of AIC and BIC requires the same sample size, we perform the comparison using an identical set of Cepheids for both models, and Model~2 remains superior. In Model~2, we find $R_2 = -0.107 \pm 0.022$, indicating that $R_G$ depends on reddening. Furthermore, Figure~\ref{fig:3} shows that the parallax residuals of Model~1 exhibit a strong dependence on reddening, whereas those of Model~2 are independent of it. These findings consistently favor Model~2, reinforcing the conclusion that $R_G$ is reddening-dependent. We therefore adopt Model~2 as our preferred calibration of the G-band PLZ relation. Recent calibrations of the 
G-band PLZ relation have also been presented by \citet{Breuval2022} and \citet{Trentin2024jan}.

To further verify the reddening dependence of $R_G$, we repeat the MCMC fitting procedure using the infrared sample in place of the \textit{Gaia} sample. This model is hereafter referred to as Model~3. The setup and parameter priors of Model~3 are the same as those of Model~2, except that the $zp$ is omitted and the likelihood is now constructed in distance space:
\begin{equation}
d_\mathrm{model} = 10^{0.2(m_G - A_G - M_G) - 2},
\end{equation}
\begin{equation}
\ln \mathcal{L} = 
-\frac{1}{2}\sum_i \left[
\frac{(d_i - d_{\mathrm{model},i})^2}{\sigma_{\mathrm{tot},i}^2}
+ \ln(2\pi\sigma_{\mathrm{tot},i}^2)
\right].
\end{equation}
The posterior distributions of the Model~3 parameters are shown in the right panel of Figure~\ref{fig:2}, and the corresponding fitting results are summarized in Table~\ref{tab:1}. All fitted parameters of Model~2 and Model~3 are consistent within $2\sigma$. For Model~3, we obtain $R_2 = -0.096 \pm 0.004$, consistent with the $R_2$ value from Model~2 within $0.5\sigma$, providing further evidence that $R_G$ is reddening-dependent.

\subsection{Distribution of $R_G$ and $R_V$ versus $E(G_{\mathrm{BP}} - G_{\mathrm{RP}})$} \label{subsec:3.3}

\begin{figure*}[ht!]
\centering
\includegraphics[width=0.49\textwidth]{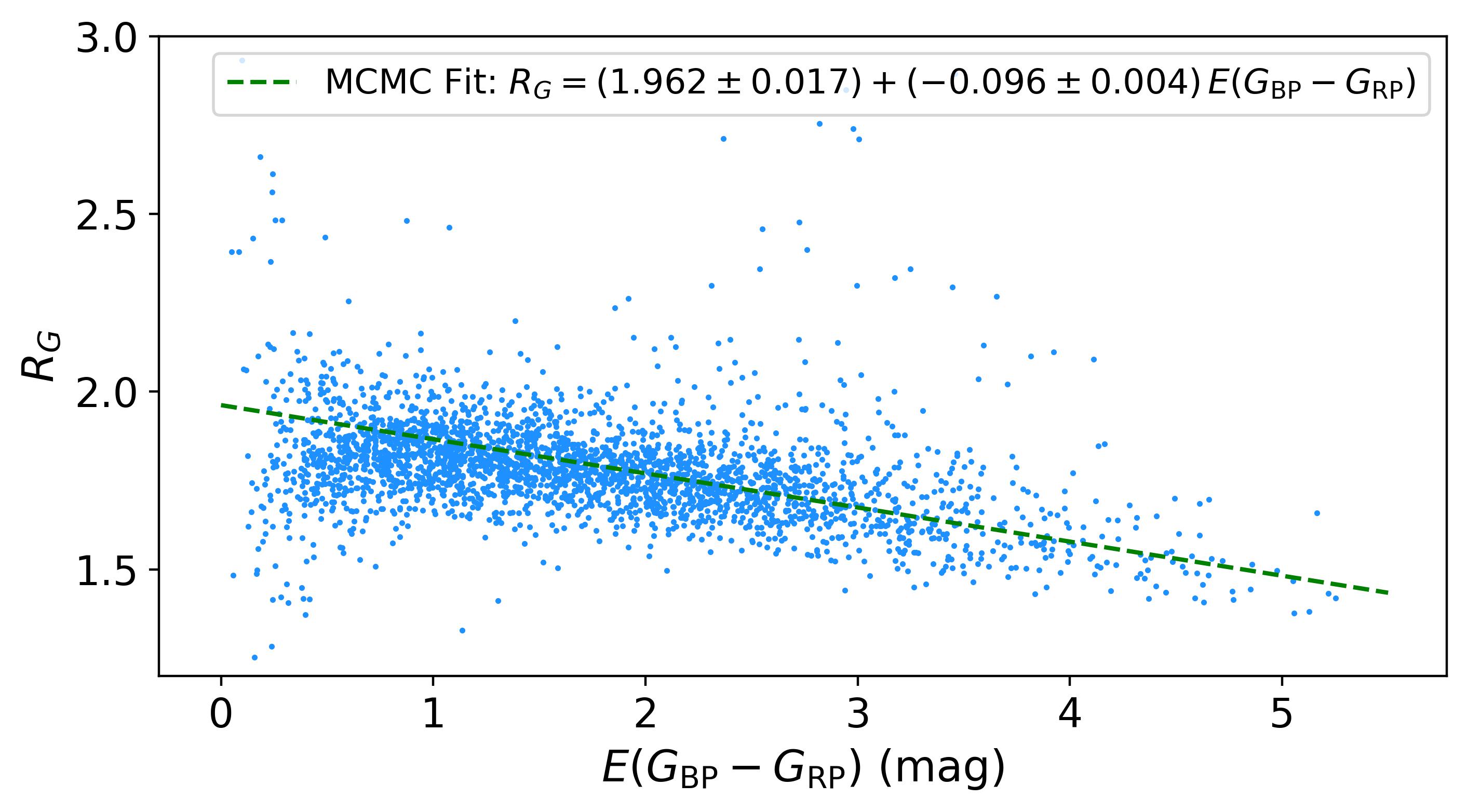}
\hfill
\includegraphics[width=0.49\textwidth]{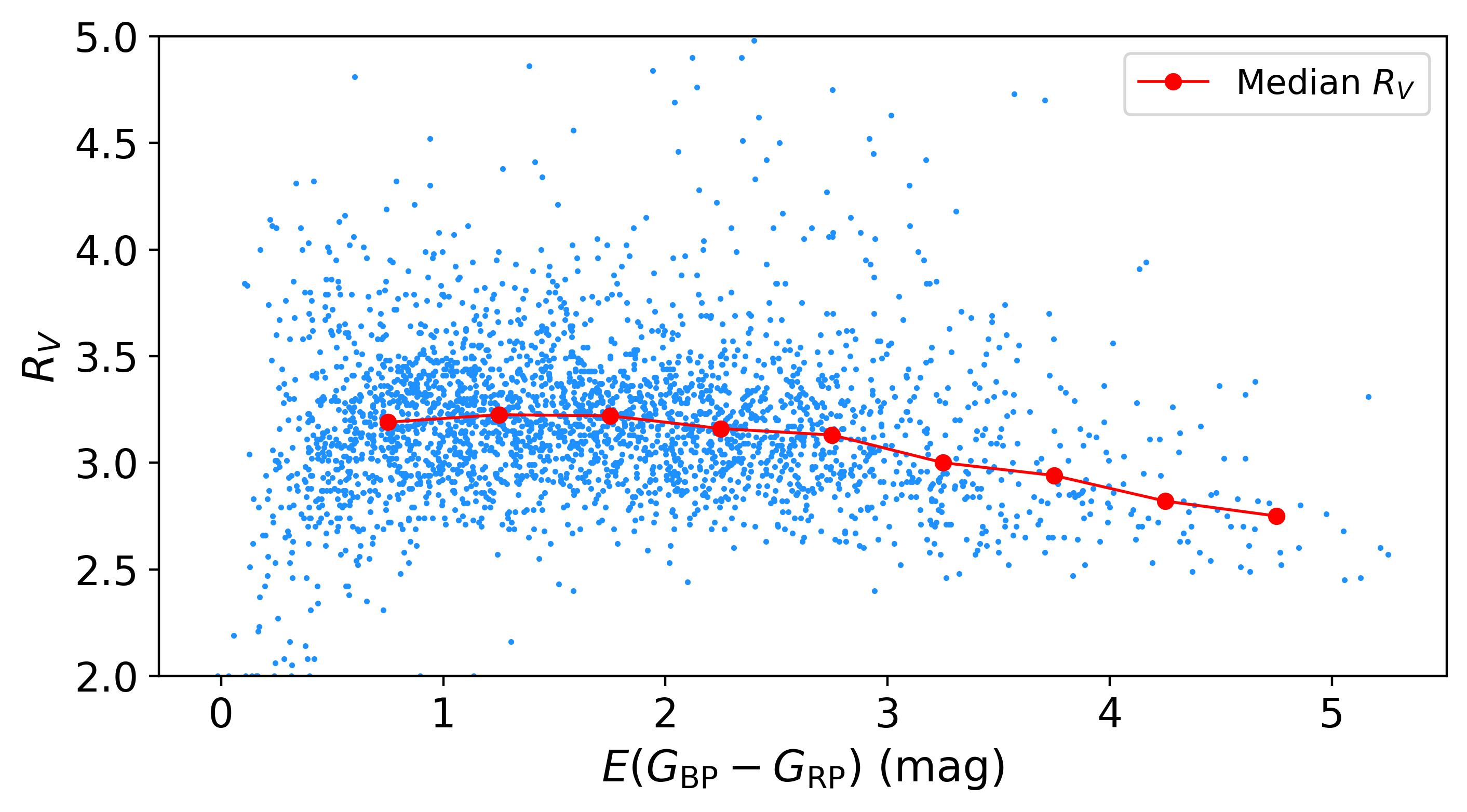}
\caption{The left panel shows the distribution of $R_G$ versus $E(G_{\mathrm{BP}} - G_{\mathrm{RP}})$ for 2,846 Cepheids, with the green dashed line representing the MCMC fitting result. The right panel presents the distribution of $R_V$ versus $E(G_{\mathrm{BP}} - G_{\mathrm{RP}})$ for the same sample, where the red points indicate the median $R_V$ values within bins of width 0.5 mag for $E(G_{\mathrm{BP}} - G_{\mathrm{RP}}) > 0.5$ mag.
\label{fig:4}}
\end{figure*}

To more intuitively illustrate the reddening dependence of $R_G$, we calculate the $R_G$ for each Cepheid using the PLZ relation calibrated with Model~3 and the infrared distances from \citet{wang2025}. The left panel of Figure \ref{fig:4} shows the distribution of $R_G$ versus $E(G_{\mathrm{BP}} - G_{\mathrm{RP}})$, with the green dashed line representing the MCMC fitting results of $R_1$ and $R_2$ from Model~3. As expected, the overall distribution of the data points aligns well with the green dashed line, except at low $E(G_{\mathrm{BP}} - G_{\mathrm{RP}})$ where mathematical effects produce larger scatter, and for a small number of outliers with unusually high $R_G$. These outliers may be caused by sample contamination or peculiar ISM.

Interestingly, the mean value of $R_2$ estimated using the convolution method is approximately $-0.046$, with an absolute value considerably smaller than the $R_2$ obtained from Model~2 and Model~3. This suggests that fixing $R_V = 3.1$ in the convolution method does not adequately capture the actual extinction behavior of Galactic Cepheids.

As shown in the right panel of Figure \ref{fig:1}, once $R_G$ and $E(G_{\mathrm{BP}} - G_{\mathrm{RP}})$ are known for each Cepheid, the convolution method can be used to derive the best-fitting $R_V$ for individual Cepheids. The right panel of Figure \ref{fig:4} displays the distribution of $R_V$ as a function of $E(G_{\mathrm{BP}} - G_{\mathrm{RP}})$, adopting the extinction law of \citet{wang2019}. Apart from points with large uncertainties at $E(G_{\mathrm{BP}} - G_{\mathrm{RP}}) < 0.5$ mag and a few outliers, $R_V$ shows a clear overall decreasing trend with increasing reddening, particularly for $E(G_{\mathrm{BP}} - G_{\mathrm{RP}}) > 2$ mag.

It is worth noting that, although adopting different extinction laws can alter the overall distribution of $R_V$, they all exhibit a similar general trend of decreasing $R_V$ with increasing reddening. A possible explanation for this trend is discussed in Section~\ref{subsec:4.2}.

\section{Discussion}\label{sec:discussion}
\subsection{Model Robustness Test} \label{subsec:4.1}

We perform several tests to assess the robustness of our analysis. First, we replace the originally adopted LMC-based PC relation with the Milky Way PC relation from \citet[see their Table 4]{Trentin2024jan} and repeat the full MCMC analysis. Except for a minor shift of approximately $1.2\sigma$ in $\alpha$, all other parameters remain essentially unchanged, indicating that our results are robust to the choice of PC relation. We also test several alternative functional forms of $R_G$, including quadratic, exponential, and logarithmic forms. None of these more complex models provides any improvement in the fit, and all models consistently recover a approximately linear decrease of $R_G$ with increasing reddening. Consequently, the linear model is selected as it offers the optimal balance between goodness-of-fit and parsimony.

When fitting the relation $
A_G = (R_1 + R_2\,E(G_{\mathrm{BP}} - G_{\mathrm{RP}}))\,E(G_{\mathrm{BP}} - G_{\mathrm{RP}})$, errors in the independent variable $E(G_{\mathrm{BP}} - G_{\mathrm{RP}})$ can introduce systematic shifts in the inferred coefficients $R_1$ and $R_2$. To quantify this purely mathematical effect, we perform a Monte Carlo simulation using the 893 Cepheids in Model~2. 
Assuming the PC relation-based $E(G_{\mathrm{BP}} - G_{\mathrm{RP}})$ values as the true reddenings and adopting $R_1=1.9$ and $R_2=-0.1$, we compute the corresponding true $A_G$, then add Gaussian noise to $E(G_{\mathrm{BP}} - G_{\mathrm{RP}})$ with a standard deviation (average value 0.082~mag) given by the quadrature sum of the photometric color dispersion and the intrinsic color dispersion. Using the perturbed $E(G_{\mathrm{BP}} - G_{\mathrm{RP}})$ and the true $A_G$, we refit $R_1$ and $R_2$ and repeat this procedure 10,000 times. The simulations yield mean recovered values of $R_1 = 1.903 \pm 0.010$ and
$R_2 = -0.105 \pm 0.005$, corresponding to relative offsets of
$\sim0.2\%$ and $\sim5\%$, respectively, with respect to the input values.
These offsets are well below the statistical uncertainties from the MCMC fit
($\sigma_{R_1} = 0.06$ and $\sigma_{R_2} = 0.022$), indicating that
this mathematical effect is negligible in our analysis. However, it could become more significant when larger reddening uncertainties are involved (e.g., when using bands with poorer photometric precision), and may be further amplified if the errors deviate from a Gaussian distribution.

\subsection{Variations of Cepheid $R_V$ in Diverse ISM} \label{subsec:4.2}

Based on the three-dimensional $R_V$ map within 5 kpc of the Sun provided by \citet{zhang2025}, combined with the infrared distances from \citet{wang2025}, we obtain $R_V$ and $A_V$ values for a total of 970 Cepheids. We find that the Cepheid $R_V$ and $A_V$ derived from this map exhibit a clear anti-correlation. However, when comparing the $R_V$ values from this map with the best-fitting $R_V$ obtained using the convolution method, we find inconsistencies. Further comparison of the $A_V$ from the map with $E(G_{\mathrm{BP}} - G_{\mathrm{RP}})$ derived from the PC relation shows that when $E(G_{\mathrm{BP}} - G_{\mathrm{RP}}) \gtrsim 1.5~\mathrm{mag}$, the map significantly underestimates the extinction. This is most likely a result of selection effects in the map at high-extinction regions, where stars with lower extinction are more easily observed.

Using a sample of 562 O-type stars, \citet{maiz2018} also reported an anti-correlation between $R_{5495}$ and $E(4405-5495)$, where 4405\,\AA{} and 5495\,\AA{} approximately correspond to the central wavelengths of the $B$ and $V$ bands. Both \citet{maiz2018} and \citet{zhang2025} found that, compared to the diffuse ISM, moderately dense translucent molecular clouds tend to exhibit lower $R_V$ (also noted by \citet{Zhang2023}), whereas H\,\textsc{ii} regions generally show higher $R_V$. However, the physical interpretations proposed by \citet{maiz2018} and \citet{Zhang2025apjl} differ.
\citet{maiz2018} attributed the variations in $R_V$ to the selective destruction of small dust grains by extreme ultraviolet (EUV) radiation \citep[e.g.,][]{Guhathakurta1989}. In this scenario, translucent molecular clouds exhibit lower $R_V$ because their high densities shield EUV photons, preserving small grains, whereas H\,\textsc{ii} regions, rich in young stars emitting EUV radiation, result in higher $R_V$ environments. In contrast, \citet{Zhang2025apjl} proposed that the lower $R_V$ observed in translucent molecular clouds is driven by an increase in the total mass of polycyclic aromatic hydrocarbons due to gas-phase accretion. Assessing which physical mechanism is more plausible requires a more detailed study, which is beyond the scope of this work.

Based on the findings from the above literature, we propose a possible explanation for the overall decreasing trend of $R_V$ with increasing reddening shown in the right panel of Figure~\ref{fig:4}. Cepheids with higher reddening preferentially trace lines of sight passing through translucent molecular clouds rather than the diffuse ISM. If translucent molecular clouds exhibit lower $R_V$ compared to the diffuse ISM, this naturally gives rise to a statistical anti-correlation between $R_V$ and reddening. Although dust grains in the dense cores of molecular clouds are expected to grow into larger grains \citep[e.g.,][]{Draine2003,Chapman2009}, the Cepheids in our sample are identified at optical bands and are therefore unlikely to lie behind such heavily extinguished regions. We also ignore H\,\textsc{ii} regions because, although they may have relatively high dust densities, their small spatial scales lead to a low probability of intersecting the lines of sight to Cepheids, and also make it difficult for them to contribute sufficiently large column densities to produce high extinction.

\subsection{Implication for Galactic Cepheid Distances} \label{subsec:4.3}

Although we find an approximately linear dependence of $R_G$ on $E(G_{\mathrm{BP}}-G_{\mathrm{RP}})$ for Galactic Cepheids, we do not recommend extrapolating distances using the parameters of Model~2. This is because the Model~2 sample is limited to the solar neighborhood and represents only the lower-extinction portion of the full Cepheid population. Consequently, the notable uncertainties in $R_1$ and $R_2$ inferred from Model~2 are significantly amplified at high extinction, leading to substantial deviations in distance estimates. We adopt $R_G$ as a tracer of reddening dependence because its approximately linear relation with $E(G_{\mathrm{BP}}-G_{\mathrm{RP}})$ in our sample allows us to construct a simple linear model that effectively disentangles non-linear effects from variations in $R_V$. However, it should be emphasized that the functional form of the reddening dependence of extinction coefficients can vary substantially depending on the bands considered, and the associated non-linear effects are likewise diverse.

As shown in Table \ref{tab:1}, the metallicity coefficients $\gamma$ are $-0.608 \pm 0.050$ for Model~1 and $-0.396 \pm 0.051$ for Model~2. This indicates that ignoring the reddening dependence of $R_G$ would lead to an overestimation of the metallicity dependence of Cepheid luminosities in the $G$-band by about $3\sigma$.
Although the reddening dependence of extinction coefficients in other bands may be less pronounced than that of $R_G$, this result serves as a cautionary note for calibrating PLZ relations in other bands. Similarly, when deriving the metallicity coefficient $\gamma$ by comparing the PL relation intercepts $\beta$ and mean metallicities between the Magellanic Clouds and the Milky Way, it is also worthwhile to account for the reddening dependence of the extinction coefficients to improve accuracy, especially within the Milky Way.

It should be noted that when an extinction coefficient exhibits a significant dependence on reddening, as in the case of $R_G$, the Wesenheit function \citep{madore1982} constructed using that coefficient becomes unsuitable for highly reddened
Cepheids, because the definition of Wesenheit magnitudes requires a fixed extinction coefficient \citep[see Appendix~D of][]{riess2022}.

This study demonstrates that both non-linear effects and the diversity of the ISM can alter extinction coefficients, thereby degrading distance accuracy. Ideally, the optimal method for determining Galactic Cepheid distances is to obtain individual $R_V$ estimates for each star while accounting for non-linear effects. However, reliable $R_V$ measurements are currently unavailable for Galactic Cepheids. Therefore, in the absence of precise $R_V$ measurements for Galactic Cepheids, we recommend using infrared-based distances, as they are less affected by non-linear effects and insensitive to $R_V$ variations caused by diverse ISM. Two viable approaches are recommended: (1) constructing Wesenheit magnitudes based on the infrared $W1$ band, for example $W_{W1,JK} = m_{W1} - R_{W1,JK} \,(m_J - m_{K_S})$ and $W_{W1,BPRP} = m_{W1} - R_{W1,BPRP} \,(m_{G_{\mathrm{BP}}} - m_{G_{\mathrm{RP}}})
$, where $R_{W1,JK} = 0.236$ and $R_{W1,BPRP} = 0.094$, adopting the extinction law of \citet[see their Table 3]{wang2019}; and (2) employing the infrared multi-band
optimal distance method to simultaneously estimate extinction and distance
\citep[e.g.,][]{Freedman1991,Chen2018,wang2025}. We have verified that these two approaches yield highly consistent distances.

It is worth noting that the Cepheid sample we use is identified in optical bands with relatively low extinction, so the non-linear effects in the infrared bands are minimal. However, for Cepheids identified in infrared bands with higher extinction, the non-linear effects on infrared extinction coefficients can still be significant and need to be taken into account \citep{Chen2018}.

\section{Conclusions}\label{sec:conclusion}

Cepheids are not only reliable standard candles but also powerful probes for studying interstellar extinction laws. Using the precise photometry and parallaxes provided by \textit{Gaia}, we find that the extinction coefficient in the $G$-band exhibits a clear anti-correlation with reddening: $R_G = 1.918 \pm 0.060 - (0.106 \pm 0.022)\,E(G_{\mathrm{BP}} - G_{\mathrm{RP}})$, a trend that is also supported by infrared-based distances. We suggest that the observed anti-correlation arises from the combined contribution of the pronounced non-linear effects inherent to the broad \textit{Gaia} bands and the diversity of the ISM: optically identified Cepheids with higher reddening tend to lie behind translucent molecular clouds with smaller average dust grains (i.e., lower $R_V$). Ignoring the reddening dependence of $R_G$ would lead to an overestimation of the metallicity dependence of Cepheid luminosities by about $3\sigma$ and a systematic underestimation of distances in highly reddened regions. The strong reddening dependence of $R_G$ also renders the Wesenheit function constructed using this coefficient inapplicable in highly reddened regions of the Galactic disk, since its definition presupposes a fixed extinction coefficient. 
Non-linear effects exist in all bands and become more pronounced in broader or shorter-wavelength bands. Therefore, until precise $R_V$ values are available for individual Galactic Cepheids, we recommend using infrared-based distances, which, compared to optical-based distances, are far less sensitive to $R_V$ variations caused by diverse ISM and, at the same reddening, are far less affected by non-linear effects.

In future work, we will attempt to determine precise $R_V$ values for each Galactic Cepheid, thereby obtaining more reliable distance estimates and investigating how $R_V$ correlates with the dust properties encountered along the line of sight.

\begin{acknowledgments}

We thank the referee for the professional comments. This work was supported by the National Natural Science Foundation of China (NSFC) through grants 12322306, 12373028, 12173047, 12233009, and 12133002. This work is supported by the China Manned Space Program with grant no. CMS-CSST-2025-A01. X.C. and S.W. acknowledge support from the Youth Innovation Promotion Association of the Chinese Academy of Sciences (CAS, Nos. 2022055 and 2023065). This work has made use of data from the European Space Agency (ESA) mission
{\it Gaia} (\url{https://www.cosmos.esa.int/gaia}), processed by the {\it Gaia}
Data Processing and Analysis Consortium (DPAC,
\url{https://www.cosmos.esa.int/web/gaia/dpac/consortium}). Funding for the DPAC
has been provided by national institutions, in particular the institutions
participating in the {\it Gaia} Multilateral Agreement.
\end{acknowledgments}

\bibliography{manuscript}{}
\bibliographystyle{aasjournalv7}



\end{document}